\documentclass[12pt]{article}
\usepackage{graphicx}
\usepackage{amsfonts,amssymb}
\usepackage{amsmath}
\usepackage{citehack,cite}

\newcommand{\non}{\nonumber \\}
\newcommand{\ve}[1]{{\bf #1}}
\newcommand{\be}{\begin{equation}}
\newcommand{\ee}{\end{equation}}
\newcommand{\bea}{\begin{eqnarray}}
\newcommand{\eea}{\end{eqnarray}}
\newcommand{\sli}{\sum\limits}
\newcommand{\ili}{\int\limits}

\newcommand{\vl}{\ve{l}}
\newcommand{\lp}{\left (}
\newcommand{\rp}{\right )}

\newcommand{\cB}{{\mathcal{B}}}
\newcommand{\cD}{{\mathcal{D}}}
\newcommand{\vk}{\ve{k}}
\newcommand{\rhok}{\rho_{\ve{k}}}
\newcommand{\rhomk}{\rho_{-\ve{k}}}

\begin{document}

\begin{center}
{\bf ANALYTICAL CALCULATION OF THE CRITICAL TEMPERATURE AND ESTIMATION OF
THE CRITICAL REGION SIZE FOR A FLUID MODEL}
\end{center}

\begin{center}
{\sc I.V. Pylyuk\footnote{e-mail: piv@icmp.lviv.ua}, M.P. Kozlovskii, O.A. Dobush}
\end{center}

\begin{center}
{\it Institute for Condensed Matter Physics
of the National Academy of Sciences of Ukraine,
1~Svientsitskii Str., 79011 Lviv, Ukraine}
\end{center}

\vspace{0.5cm}

{\small
An analytical procedure for calculating the critical temperature and estimating
the size of critical region for a cell fluid model is developed. Our numerical
calculations are illustrated by the case of the Morse potential parameters
characterizing the alkali metals (sodium and potassium). The critical temperatures
found for liquid sodium and potassium as solutions of the resulting quadratic
equation agree with experimental data. The expression for the relative temperature
determining the critical region size is obtained proceeding from
the condition for the critical regime existence. In the cases of sodium and
potassium, the value of this temperature is of the order of a few hundredths.
}

\vspace{0.5cm}

PACS numbers: 05.70.Ce, 64.60.F-, 64.70.F-

Keywords: cell fluid model, Morse interaction potential, grand partition function,
recurrence relations, critical temperature, critical region

\section{Introduction}
\label{sec:1}

Critical phenomena in simple and multicomponent liquid systems have been
the subject of many theoretical and experimental studies during
the past decades (see, for example,
\cite{smch183,p104,bul2,vch106,y115,tk116,cs118,yl118,v118,pyy121,gm122}).
These systems are of great practical importance as well as very interesting from
theoretical point of view. Experimental work is the basis for having a database
of properties of pure fluids and mixtures, and theoretical models can provide
a large amount of information of a fluid in a rapid, clean and cheap manner.

This paper supplements our previous study \cite{pkdd123} based on a cell
fluid model. Interaction in the system is chosen in the form of the Morse
potential possessing the Fourier transform. Despite the great successes in
the investigation of Morse fluids made by means of various methods
(for example, the $NpT$ plus test particle method
\cite{okumura_00}, the grand-canonical transition matrix Monte Carlo
method \cite{singh}, the integral equations approach \cite{apf_11},
molecular dynamics simulations in a canonical ensemble \cite{martinez}),
the statistical description of the behavior of the mentioned fluids near
the critical point on the microscopic level without any general assumptions
are still of interest.

In \cite{pkdd123}, the cell fluid model is used for studying the behavior
of a simple Morse fluid in the immediate vicinity of the liquid--gas critical point.
The parameters of the Morse interaction potential used for calculations
are inherent to alkali metals (sodium and potassium). The values of the critical
temperature and the size of the critical region are given in \cite{pkdd123}
without describing the method for obtaining them. In the present paper,
we describe an analytical procedure for calculating the critical temperature
and estimating the critical region size for a fluid model.

\section{Model for Describing a Fluid System}
\label{sec:2}

The volume of the system $V$ composed of $N$ interacting particles is
conventionally divided into $N_v$ cells, each of volume $v=V/N_v=c^3$ ($c$ is
the linear size of a cubic cell) \cite{kpd118,p120,pkdd123}. Note that,
in contrast to a cell gas model (where it is assumed that a cell may contain
only one particle or does not contain any particle) \cite{rebenko_13,rebenko_15},
a cell within our approach may contain more than one particle \cite{pd120,pk722}.

The grand partition function of the cell fluid model within the framework
of the grand canonical ensemble has the form \cite{kpd118,p120,pkdd123}
\be
\Xi \! = \! \sli_{N=0}^{\infty} \!\! \frac{(z)^N}{N!} \! \int \limits_{V} \!\! (dx)^N
\! \exp \! \left[-\frac{\beta}{2} \sli_{\vl_1,\vl_2\in\Lambda} \!\!\!
\tilde U_{l_{12}} \rho_{\vl_1} (\eta) \rho_{\vl_2} (\eta) \right],
\label{0d1fb2}
\ee
where $z = e^{\beta \mu}$ is the activity, $\beta=1/(kT)$
is the inverse temperature, and $\mu$ is the chemical potential.
Integration with respect to coordinates of all the particles
$x_i = (x_{i}^{(1)},x_{i}^{(2)}, x_{i}^{(3)})$ is noted as
$\int \limits_{V} (dx)^N = \int \limits_{V} dx_1 \cdots \int \limits_{V} dx_N$,
and $\eta = \{ x_1 , \ldots , x_N \}$ is the set of coordinates.
The interaction potential $\tilde U_{l_{12}}$ is a function of the distance
$l_{12}= |\ve{l}_{1}-\ve{l}_{2}|$ between cells. Each vector $\ve{l}_i$ belongs
to the set
\be
\Lambda = \Big\{ \vl = (l_1, l_2, l_3)|l_i = c m_i;
m_i=1,2,\ldots,N_a; i=1,2,3; N_v=N_a^3 \Big\}.
\label{jml2020-1fb2}
\ee
Here, $N_a$ is the number of cells along each axis. The occupation numbers of cells
appearing in Eq. (\ref{0d1fb2}) are defined as
\be
\rho_{\vl}(\eta) = \sli_{x \in \eta} I_{\Delta_{\vl}(x)},
\label{0d2fb2}
\ee
where $I_{\Delta_{\vl}(x)}$ is the indicators of cubic cells
$\Delta_{\vl} = (-c/2,c/2]^3 \subset \mathbb{R}^3$,
that is, $I_{\Delta_{\vl}(x)}=1$ if $x\in \Delta_{\vl}$ and
$I_{\Delta_{\vl}(x)}=0$ otherwise.
The role of the interaction potential $\tilde U_{l_{12}}$ is played by
the Morse potential:
\bea
&&
\tilde U_{l_{12}} = \Psi_{l_{12}} - U_{l_{12}}; \non
&&
\Psi_{l_{12}} = D e^{-2(l_{12}- 1)/\alpha_R}, \non
&&
U_{l_{12}} = 2 D e^{-(l_{12}- 1)/\alpha_R}.
\label{0d4fb2}
\eea
Here, $\Psi_{l_{12}}$ and $U_{l_{12}}$ are the repulsive and attractive parts
of the potential, respectively, and $\alpha_R = \alpha/R_0$ ($\alpha$ is
the effective interaction radius). The parameter $R_0$ corresponds
to the minimum of the function $\tilde U_{l_{12}}$, and $D$ determines
the depth of a potential well. Note that the $R_0$-units are used for
length measuring in terms of convenience. As a result,
$R_0$- and $R_0^3$-units are used for the linear size of each cell $c$ and
volume $v$, respectively.

\section{Grand Partition Function, Recurrence Relations and Their Solutions}
\label{sec:3}

When calculating the grand partition function, we use the method of
``layer-by-layer'' integration with respect to collective variables (CV)
$\rho_{\bf k}$ proposed by Yukhnovskii for magnetic
systems \cite{ymo287,YuKP_2001,ykp202,kpp406,kpp606}. This procedure
has already been represented for the simpler $\rho^4$ model in \cite{kpd118}.
As a result of step-by-step calculation of the grand
partition function, the number of integration variables in the expression
for this quantity decreases gradually. The grand partition function of
the cell fluid is then represented as a product of the partial partition
functions of individual layers and the integral of the ``smoothed'' effective
measure density:
\be
\Xi  = 2^{(N_{n+1}-1)/2}G_\mu (Q(r_0))^{N_v} Q_1 ...Q_n [Q(P_n)]^{N_n+1}
\int W_{n+1}(\rho) (d\rho)^{N_{n+1}}.
\label{1d31fa2}
\ee
The quantities $G_\mu$ and $Q(r_0)$ are given in \cite{kpd118}, $n$ is the layer number
in the CV phase space, $N_{n+1} = N_vs^{-3(n+1)}$,
$s$ is the parameter of division of the CV phase space into layers. The partial
partition function of the $n$th layer
\be
Q_n = \left[ Q(P_{n-1}) Q(d_n)\right]^{N_n}
\label{1d33fa2}
\ee
is expressed by the quantities
\begin{align}
& Q(d_n) = (2\pi)^{1/2} \lp \frac{3}{a_4^{(n)}}\rp^{1/4} \exp\lp \frac{x^2_n}{4}\rp U(0,x_n), \non
& Q(P_n) = (2\pi)^{-1/2} \lp \frac{a_4^{(n)}}{\varphi(x_n)}\rp^{1/4} s^{3/4} \exp\lp \frac{y^2_n}{4}\rp
U(0,y_n).
\label{1d32fa2}
\end{align}
The variable $y_n = s^{3/2} U(x_n) (3/\varphi(x_n))^{1/2}$ is a function of
the variable $x_n=g_n(B_{n+1},B_n) (3/a_4^{(n)})^{1/2}$. This variable $x_n$ is
determined by the coefficients $g_n(k)$ and $a_4^{(n)}$ appearing in the expression
for the non-Gaussian quartic density of measure of the $n$th layer. The special
functions
\be
U(x_n) = U(1,x_n) / U(0,x_n)
\label{cmp2018-1fa2}
\ee
and
\be
\varphi(x_n) = 3U^2 (x_n) + 2x_n U(x_n) - 2
\label{cmp2018-2fa2}
\ee
are combinations of the parabolic cylinder functions
\be
U(a,t) = \frac{2}{\Gamma(a+\frac{1}{2})} e^{-t^2/4} \ili_0^\infty x^{2a} \exp (-tx^2 - \frac{1}{2} x^4) dx.
\label{cmp2018-3fa2}
\ee
The argument $t$ may be the main variable $x_n$ or the intermediate variable $y_n$.
The effective quartic measure density $ W_{n+1}(\rho)$
appearing in Eq. (\ref{1d31fa2}) has the form
\bea
W_{n_p+1}(\rho) & = & \exp \Biggl[ a_1^{(n+1)} \sqrt{N_{n+1}} \rho_0 -
\frac{1}{2} \sli_{\vk\in\cB_{n+1}} g_{n+1} (k) \rhok\rhomk \non
&&
- \frac{a_4^{(n+1)}}{4!} \frac{1}{N_{n+1}} \mathop{\sum_{\vk_1,\ldots,\vk_4}}\limits_{\vk_i\in\cB_{n+1}}
\rho_{\vk_1}\cdots\rho_{\vk_4} \delta_{\vk_1+\cdots+\vk_4} \Biggr],
\label{cmp2018-4fa2}
\eea
where $\delta_{\vk_1+\cdots+\vk_4}$ is the Kronecker symbol, and the region of
wave vectors $\vk$ is defined as
\bea
\cB_{n+1} & = & \Big\{ \vk = (k_1, k_2, k_3) | k_i = - \frac{\pi}{c_{n+1}} +
\frac{2\pi}{c_{n+1}} \frac{n_i}{N_{n+1,i}};
n_i \! = \! 1,2,\ldots,N_{n+1,i}; \non
&&
i \! = \! 1,2,3; N_{n+1} \! = \! N_{n+1,1}^3 \Big\}.
\label{cmp2018-5fa2}
\eea

The coefficients in the exponent of the quartic measure densities of the $(n+1)$th
and $n$th layers are connected through the general recurrence
relations (RR) \cite{kpd118}
\bea
&&
w_{n+1} = s^{\frac{d+2}{2}} w_n, \non
&&
r_{n+1} = s^2 [-q + (r_n+q) N(x_n)], \non
&&
u_{n+1} = s^{4-d} u_n E(x_n).
\label{1d35fa2}
\eea
with the initial conditions
\bea
&&
w_0=M(\beta W(0))^{1/2}, \quad r_0 = 1 - \beta W(0) \tilde a_2, \non
&&
u_0 = a_4 (\beta W(0))^2.
\label{1d11fa2}
\eea
Here,
\bea
&&
w_{n+1} =s^{(n+1)} a_1^{(n+1)}, \quad r_{n+1} = s^{2(n+1)} g_{n+1}(0), \non
&&
u_{n+1} = s^{4(n+1)} a_4^{(n+1)}.
\label{cmp2018-6fa2}
\eea
The quantity $q$ is associated with the averaging of the wave vector square,
$d = 3$ is the space dimension. The functions $N(x_n)$ and $E(x_n)$ satisfy the
expressions
\be
N(x_n) = \frac{y_n U(y_n)}{x_n U(x_n)}, \quad
E(x_n) = s^{2d} \frac{\varphi(y_n)}{\varphi(x_n)}.
\label{cmp2018-7fa2}
\ee
The quantity $M$ is expressed by the chemical potential, $W(0)$ is the Fourier
transform of the effective interaction potential at zero value of the wave vector,
$\tilde a_2$ and $a_4$ is the coefficients in the initial expression for the
grand partition function (see \cite{kpd118}).

The coordinates of the fixed point ($w^*$, $r^*$, $u^*$) can be found from the conditions
\be
w_n \! = \! w_{n+1} \! = \! w^*, \quad r_n \! = \! r_{n+1} \! = \! r^*,
\quad u_n \! = \! u_{n+1} \! = \! u^*.
\label{cmp2018-8fa2}
\ee
For $w^*$, we have $w^*=0$, since $s>1$. The third equation for $u_{n+1}$
[see Eqs. (\ref{1d35fa2})] yields the relation
\be
s E(x^*) = 1,
\label{1d36fa2}
\ee
which juxtaposes own $x^*$ to each $s$. Our calculations are performed for
some fixed value of the parameter $s=s^*=3.5977$. For such a preferred value of $s$
nullifying the variable $x_n=(r_n +q) (3/u_n)^{1/2}$ at the fixed point ($x^*=0$),
the mathematical description becomes less complicated. Using the second equation
for $r_{n+1}$ [see Eqs. (\ref{1d35fa2})], we arrive at the following expression:
\be
(u^*)^{1/2}  = q(1-s^{-2}) \sqrt 3 U(x^*) / (y^* U(y^*)).
\label{1d37fa2}
\ee
Thus, the fixed point coordinates are $w^*=0$, $r^*=-q$, and $u^*$ is determined
from Eq. (\ref{1d37fa2}). Note that the variable $y_n$ takes large values.
Taking this into account, we obtain the expressions
\bea
&&
w_{n+1} = s^{\frac{d+2}{2}} w_n, \non
&&
r_{n+1} = s^2 \left[ - q + \frac{\sqrt{u_n}}{\sqrt{3}} \frac{1}{U(x_n)} -
\frac{1}{2s^3} \frac{\sqrt{u_n}}{\sqrt{3}} \frac{\varphi(x_n)}{U^3(x_n)}\right], \non
&&
u_{n+1} = s u_n \frac{\varphi(x_n)}{3 U^4(x_n)} \left[ 1 -
\frac{7}{2} s^{-3} \frac{\varphi(x_n)}{U^2(x_n)}\right]
\label{cmp2018-9fa2}
\eea
and
\be
(u^*)^{1/2} = q(1-s^{-2}) \sqrt{3} U(x^*) \left[ 1 + \frac{3}{2} (y^*)^{-2} \right]
\label{cmp2018-10fa2}
\ee
corresponding to the relations (\ref{1d35fa2}) and (\ref{1d37fa2}), respectively.
The quantity $q$ does not depend on temperature, so $r^*$ and $u^*$ are also
not functions of temperature. They depend on $\alpha_R = \alpha/R_0$.

The solutions of RR in the vicinity of the fixed point ($w^*$, $r^*$, $u^*$)
can be written through the eigenvalues of the matrix $\cal R$ of the linear
transformation
\be
\left(
\begin{array}{lll}
w_{n+1}-w^* \\
r_{n+1}-r^* \\
u_{n+1}-u^*
\end{array}
\right) = \cal R
\left(
\begin{array}{lll}
w_{n}-w^* \\
r_{n}-r^* \\
u_{n}-u^*
\end{array}
\right).
\label{1d38fa2}
\ee
They assume the form \cite{kpd118}
\bea
&&
w_n = w_0 E_1^n, \non
&&
r_n = r^* + c_1 E_2^n + c_2 R E_3^n, \non
&&
u_n = u^* + c_1 R_1 E_2^n + c_2  E_3^n,
\label{1d40fa2}
\eea
where $E_l$ are the eigenvalues of the matrix $\cal R$.
The coefficients
\bea
&&
c_1 = [r_0 - r^* + (u^* - u_0) R] \cD^{-1}, \non
&&
c_2 = [u_0 - u^* + (r^* - r_0) R_1] \cD^{-1}
\label{1d41fa2}
\eea
are determined by the eigenvalues and elements of the renormalization group
linear transformation matrix, coordinates of the fixed point,
and initial coefficients $\tilde a_2$, $a_4$. The quantities $R$, $R_1$, and $\cD$
appearing in Eqs. (\ref{1d40fa2}) and (\ref{1d41fa2}) satisfy
the expressions
\bea
&&
R = R^{(0)}(u^*)^{-1/2}, \quad R^{(0)} = \frac{R_{23}^{(0)}}{E_3-R_{22}}, \non
&&
R_1 = R_1^{(0)}(u^*)^{1/2}, \quad R_1^{(0)} = \frac{E_2-R_{22}}{R_{23}^{(0)}}, \non
&&
\cD = \frac{E_2-E_3}{R_{22}-E_3}.
\label{1d42fa2}
\eea
In the case when $s=s^*$, we get the following numerical values:
\bea
&&
E_1 = s^{\frac{d+2}{2}} = 24.551, \quad E_2 = 8.308, \quad E_3 = 0.374, \non
&&
R^{(0)} = -0.530, \quad R_1^{(0)} = 0.162, \quad \cD = 1.086.
\label{cmp2018-11fa2}
\eea

Let us represent $c_1(T)$ and $c_2(T)$ from Eqs. (\ref{1d41fa2}) as expansions
in powers of the relative temperature $\tau=(T-T_c)/T_c$ ($T_c$ is the critical
temperature). Using the expressions (\ref{1d11fa2}) for $r_0$ and $u_0$ and
taking into account the fact that the coordinates of the fixed point of
RR (\ref{1d35fa2}) are not functions of temperature, we can write
\bea
&&
c_1 = c_{10} + c_{11}\tau + c_{12}\tau^2, \non
&&
c_2 = c_{20} + c_{21}\tau + c_{22}\tau^2.
\label{1d45fa2}
\eea
Here, $c_{10}=0$, because of the equation $c_1(T_c)=0$, which, actually, is used
to determine the critical temperature. Other coefficients in the expression
for $c_1$ are defined as
\bea
&&
c_{11} \!\! = \!\! \beta_c W(0) \cD^{-1} \!\! \left[ \! \tilde a_2 \! + \! 2R^{(0)} \beta_c W(0) a_4 (u^*)^{-1/2} \!\right], \non
&&
c_{12} \!\! = \!\! - \beta_c W(0) \cD^{-1} \!\! \left[ \! \tilde a_2 \! + \! 3R^{(0)} \beta_c W(0) a_4 (u^*)^{-1/2} \!\right].
\label{1d46fa2}
\eea
For the coefficients $c_{2l} (l=0,1,2)$, we find
\bea
&&
c_{20} \!\! = \!\! \cD^{-1} \!\! \Big[ \! - \! u^* \! - \! R_1^{(0)} \sqrt{u^*} (1 \! + \! q) \! + \!
R_1^{(0)} \sqrt{u^*} \tilde a_2 \beta_c W(0) + a_4 (\beta_c W(0))^2 \!\Big], \non
&&
c_{21} = - \cD^{-1} \Big[ R_1^{(0)} \sqrt{u^*} \tilde a_2 \beta_c W(0) + 2a_4 (\beta_c W(0))^2 \Big], \non
&&
c_{22} =  \cD^{-1} \Big[ R_1^{(0)} \sqrt{u^*} \tilde a_2 \beta_c W(0) \! + \! 3a_4 (\beta_c W(0))^2 \Big].
\label{1d47fa2}
\eea

Let us now proceed to the calculation of the critical temperature and
the estimation of the size of the critical region.

\section{Critical Temperature and Critical Region Size}
\label{sec:4}

There is a temperature $T=T_c$ at which
\be
c_1(T_c) = 0.
\label{1d43fa2}
\ee
When $M=0$ and $T=T_c$, all three quantities $w_n$, $r_n$, and $u_n$
from Eqs. (\ref{1d40fa2})  go to their fixed values at $n\rightarrow \infty$.
Taking into account the expression for $c_1$ [see Eqs. (\ref{1d41fa2})],
we can rewrite the equation (\ref{1d43fa2}) for the critical
temperature $T_c$ in the following form:
\be
1 - \tilde a_2\beta_c W(0) - r^* - R(a_4(\beta_c W(0))^2 - u^*) = 0.
\label{cmp2018-12fa2}
\ee
Since $r^*=-q$, we obtain the equation
\be
1 + q + R^{(0)} \sqrt{u^*} - \tilde a_2 \beta_c W(0) - R^{(0)} \frac{a_4}{\sqrt{u^*}} (\beta_c W(0))^2 = 0,
\label{1d44fa2}
\ee
where $\beta_c=1/(kT_c)$, and the value of $R^{(0)}$ is given in
Eqs. (\ref{cmp2018-11fa2}). This equation allows us to find the critical
temperature of the fluid model as a function of microscopic parameters of
the interaction potential and coordinates of the fixed point of RR.
The calculations in this paper are performed for the parameters of the Morse
interaction potential taken from \cite{kpd118,p120}, which correspond to the data
for sodium and potassium \cite{singh}. We have $R_0/\alpha = 2.9544$ for sodium (Na)
and $R_0/\alpha = 3.0564$ for potassium (K).

The quantities included in equation (\ref{1d44fa2}) and the critical temperatures
obtained for liquid metals (Na and K) from this equation are given in
Table~\ref{tab_1_ujp23_pkd_2}. Numerical values of the critical temperature
represented in the form of reduced dimensionless units are obtained in different ways:
from our present researches on the basis of the cell fluid model (see $kT_c/D$
in Table~\ref{tab_1_ujp23_pkd_2}), from Monte Carlo simulation results for
the continuous system with the Morse potential in the grand canonical ensemble
(see $kT_c/D$ \cite{singh}), and from experiment (see $kT_c/D$ \cite{hensel}).
\begin{table*}[htbp]
\noindent\caption{The quantities appearing in equation (\ref{1d44fa2}) for
the critical temperature and the values of the dimensionless
critical temperature for liquid alkali metals (Na and K). The constant $D$
is the energy parameter for the Morse potential ($D=0.9241\times 10^{-13}$ ergs
for Na and $D=0.8530\times 10^{-13}$ ergs
for K \cite{singh}).}\vskip3mm\tabcolsep4.5pt
\label{tab_1_ujp23_pkd_2}

\noindent{\footnotesize\begin{tabular}{|c|c|c|c|c|c|c|c|c|}
 \hline \multicolumn{1}{|c}
{\rule{0pt}{5mm}Metal} & \multicolumn{1}{|c}{$q$}&
\multicolumn{1}{|c}{$u^*$}&
\multicolumn{1}{|c}{$\tilde a_2$}&
\multicolumn{1}{|c}{$a_4$}&
\multicolumn{1}{|c|}{$W(0)/D$}&
\multicolumn{1}{|c}{$kT_c/D$}&
\multicolumn{1}{|c}{$kT_c/D$ \cite{singh}}&
\multicolumn{1}{|c|}{$kT_c/D$ \cite{hensel}}\\[2mm]%
\hline%
\rule{0pt}{5mm}Na& 1.236& 3.626& 0.324& 0.038& 17.769& 4.028& 5.874& 3.713 \\%
K& 0.880& 1.839& 0.313& 0.039& 16.072& 3.304& 5.050& 3.690 \\[2mm]%
\hline
\end{tabular}}
\end{table*}
As can be seen from Table~\ref{tab_1_ujp23_pkd_2}, our estimates of the critical
temperature for Na and K agree better with the experimental data \cite{hensel}
than the numerical results \cite{singh} obtained by Monte Carlo simulations.

The renormalization group symmetry that occurs in the system indicates a change
in the temperature behavior of the thermodynamic functions when the temperature
approaches $T_c$. The absence of the region of the critical regime means that
the system will be described by a Gaussian regime of fluctuations, which leads
to classical values of critical exponents. From the point of view of
the theoretical description of the phase transition at the microscopic level,
the critical exponents are completely determined by the critical regime region.
The transition of classical critical exponents (the region of Gaussian
fluctuations of the order parameter) to non-classical ones is determined by
the quantity $\tau^*$ and takes place only if there is the critical
regime of fluctuations. This quantity determines the size of the critical
region.

The size of the critical region are an important element of every theoretical
scheme describing the phase transition. The Ginzburg criterion for determining
the size of the critical region of temperatures is well known (see, for example,
\cite{llmo176,lgmo177} and references cited herein). In this paper,
an alternative option is described.

Let us estimate the order of magnitude of $\tau^*$.
The solutions of the renormalization group type (\ref{1d40fa2}) correspond to
the region of the critical regime. In these solutions, the terms proportional
to $E_3^n$ describe the entry to the critical regime, and the terms
proportional to $E_2^n$ describe the exit from the critical regime.
The condition for the critical regime existence is that the exit
from the critical regime for $n \rightarrow 1$ should not prevail over
the entry to this regime. Using the solutions (\ref{1d40fa2}) and this
condition, we can determine the temperature range $\tau < \tau^*$ in which
the critical regime occurs. The temperature $\tau^*$ will be equal to
the magnitude (the absolute value) of the smallest root ($\tau_1^*$
or $\tau_2^*$) of the two equations
\bea
&&
c_1(\tau_1^*) E_2 = c_2(\tau_1^*) R E_3, \non
&&
c_1(\tau_2^*) R_1 E_2 = c_2(\tau_2^*) E_3.
\label{f1_ujp23_pkd_2}
\eea
Equations (\ref{f1_ujp23_pkd_2}) accurate to within $\tau^*$ assume
the following form:
\bea
&&
c_{11} \tau_1^* E_2 = (c_{20} + c_{21} \tau_1^*) \frac{R^{(0)}}{\sqrt{u^*}} E_3, \non
&&
c_{11} \tau_2^* R_1^{(0)} \sqrt{u^*} E_2 = (c_{20} + c_{21} \tau_2^*) E_3.
\label{f2_ujp23_pkd_2}
\eea
The first and second equations (\ref{f2_ujp23_pkd_2}) have the solutions
\be
\tau_1^* = \frac{c_{20} \frac{R^{(0)}}{\sqrt{u^*}} E_3}{c_{11} E_2 - c_{21} \frac{R^{(0)}}{\sqrt{u^*}} E_3}
\label{f3_ujp23_pkd_2}
\ee
and
\be
\tau_2^* = \frac{c_{20} E_3}{c_{11} R_1^{(0)} \sqrt{u^*} E_2 - c_{21} E_3},
\label{f4_ujp23_pkd_2}
\ee
respectively. The quantities $R^{(0)}$, $R_1^{(0)}$, $E_2$, and $E_3$
\begin{table}[htbp]
\noindent\caption{The coefficients $c_{11}$, $c_{20}$, and $c_{21}$
appearing in Eqs. (\ref{f3_ujp23_pkd_2}) and (\ref{f4_ujp23_pkd_2})
and the values of the relative temperatures $\tau_1^*$, $\tau_2^*$,
and $\tau^*$ obtained for Na and K.}\vskip3mm\tabcolsep4.5pt
\label{tab_2_ujp23_pkd_2}

\noindent{\footnotesize\begin{tabular}{|c|c|c|c|c|c|c|}
 \hline \multicolumn{1}{|c}
{\rule{0pt}{5mm}Metal} & \multicolumn{1}{|c}{$c_{11}$}&
\multicolumn{1}{|c}{$c_{20}$}&
\multicolumn{1}{|c}{$c_{21}$}&
\multicolumn{1}{|c}{$\tau_1^*$}&
\multicolumn{1}{|c|}{$\tau_2^*$}&
\multicolumn{1}{|c|}{$\tau^*$}\\[2mm]%
\hline%
\rule{0pt}{5mm}Na& 0.942& $-2.894$& $-1.755$& 0.039& $-0.352$& 0.04 \\%
K& 0.735& $-0.910$& $-2.020$& 0.023& $-0.162$& 0.02 \\[2mm]%
\hline
\end{tabular}}
\end{table}
are presented in Eqs. (\ref{1d42fa2}) and (\ref{cmp2018-11fa2}). The values
of the fixed point coordinate $u^*$ for Na and K are given in
Table~\ref{tab_1_ujp23_pkd_2}. Table~\ref{tab_2_ujp23_pkd_2} contains
numerical estimates for $c_{11}$, $c_{20}$, and $c_{21}$ as well as for
$\tau_1^*$, $\tau_2^*$, and $\tau^*$.
Thus, we obtain $\tau^* = \tau_1^* \approx 0.04$ (in the case of liquid
sodium) and $\tau^* = \tau_1^* \approx 0.02$ (in the case of potassium).

\section{Conclusions}
\label{sec:5}

A calculation technique for estimating the critical temperature and the size
of the critical region for a fluid system is elaborated within the cell
fluid model framework. For this purpose, we use the expressions (\ref{1d40fa2})
for solutions of recurrence relations between the coefficients of the effective
measure densities. The solutions (\ref{1d40fa2}) have the general form of
the renormalization group solutions obtained by Wilson
(see, for example, \cite{wk174}) and differ from them by
explicit expressions for $c_1$ and $c_2$.

In this paper, the calculations are performed for the Morse potential parameters
characterizing real substances (sodium and potassium metals).

The equation for the critical temperature is obtained.
The critical temperature is calculated and not introduced into the Hamiltonian
of the system phenomenologically, as is done in the field theory approach or
in the Landau theory. In the Landau theory, the quantity $T - T_c$ is included
in the coefficient of the second power of the order parameter.

Our values of the critical temperature for liquid alkali metals (Na and K)
agree more closely with the experimental data \cite{hensel} than Monte Carlo
simulation results from \cite{singh}.

The expression (\ref{f3_ujp23_pkd_2}) makes it possible to find the value
of the temperature $\tau^*$ at which the coordinate of the point of entry to
the critical regime coincides with the coordinate of the point of exit from it.
This means that there is no region of the critical regime for
the temperature range $\tau > \tau^*$, but such a region exists for
the temperature range $\tau < \tau^*$. The value of the temperature $\tau^*$
determining the critical region size is of the order of a few hundredths
($\tau^* = 0.04$ in the case of liquid sodium and $\tau^* = 0.02$ for potassium).
The region of interest for most applications of supercritical fluids
covers this temperature value (usually $1 < T/T_c < 1.1$
(or $0 < \tau < 0.1$) \cite{ekd196}).


\begin{thebibliography}{32}

\bibitem{smch183} J.M.H.~Levelt Sengers, G.~Morrison, R.F.~Chang.
Critical behavior in fluids and fluid mixtures.
\textit{Fluid Phase Equilib.} {\bf 14}, 19 (1983),
https://doi.org/10.1016/0378-3812(83)80113-7.

\bibitem{p104} S.~Pittois, B.~Van Roie, C.~Glorieux, J.~Thoen.
Thermal conductivity, thermal effusivity, and specific heat capacity
near the lower critical point of the binary liquid mixture
n-butoxyethanol--water. \textit{J.Chem. Phys.} {\bf 121}, 1866 (2004),
https://doi.org/10.1063/1.1765652.

\bibitem{bul2}
Y.B.~Melnichenko, G.D.~Wignall, D.R.~Cole, H.~Frielinghaus, L.A.~Bu\-la\-vin.
Liquid-gas critical phenomena under confinement: small-angle neutron
scattering studies of CO$_2$ in aerogel. \textit{J. Mol. Liq.}
{\bf 120}, 7 (2005),
https://doi.org/10.1016/j.molliq.2004.07.070.

\bibitem{vch106} A.N.~Vasil'ev, A.V.~Chalyi.
Critical parameters and pair correlations in confined multicomponent
liquids. \textit{Condens. Matter Phys.} {\bf 9}, 65 (2006),
https://doi.org/10.5488/CMP.9.1.65.

\bibitem{y115} I.R.~Yukhnovskii.
Phase transitions in a vicinity of the vapor-liquid critical point.
\textit{Ukr. J. Phys. Reviews} {\bf 10}, 33 (2015) [in Ukrainian],
https://ujp.bitp.kiev.ua/index.php/ujp/article/view/2019662.

\bibitem{tk116} I.~Tsivintzelis, G.M.~Kontogeorgis.
Modelling phase equilibria for acid gas mixtures using the CPA equation of state.
Part VI. Multicomponent mixtures with glycols relevant to oil and gas and
to liquid or supercritical CO$_2$ transport applications.
\textit{J. Chem. Thermodyn.} {\bf 93}, 305 (2016),
https://doi.org/10.1016/j.jct.2015.07.003.

\bibitem{cs118} P.~de Castro, P.~Sollich.
Critical phase behavior in multi-component fluid mixtures: Complete
scaling analysis. \textit{J. Chem. Phys.} {\bf 149}, 204902 (2018),
https://doi.org/10.1063/1.5058719.

\bibitem{yl118} T.J.~Yoon, Y.-W.~Lee.
Current theoretical opinions and perspectives on the fundamental
description of supercritical fluids. \textit{J. Supercrit. Fluids}
{\bf 134}, 21 (2018),
https://doi.org/10.1016/j.supflu.2017.11.022.

\bibitem{v118} L.F.~Vega.
Perspectives on molecular modeling of supercritical fluids: From
equations of state to molecular simulations. Recent advances,
remaining challenges and opportunities. \textit{J. Supercrit. Fluids}
{\bf 134}, 41 (2018),
https://doi.org/10.1016/j.supflu.2017.12.025.

\bibitem{pyy121} Y.X.~Pang, M.~Yew, Y.~Yan {\em et al.} Application of
supercritical fluid in the synthesis of graphene materials: a review.
\textit{J. Nanopart. Res.} {\bf 23}, 204 (2021),
https://doi.org/10.1007/s11051-021-05254-w.

\bibitem{gm122} I.R.~Graf, B.B.~Machta.
Thermodynamic stability and critical points in multicomponent mixtures
with structured interactions. \textit{Phys. Rev. Res.}
{\bf 4}, 033144 (2022),
https://doi.org/10.1103/PhysRevResearch.4.033144.

\bibitem{pkdd123} I.V.~Pylyuk, M.P.~Kozlovskii, O.A.~Dobush, M.V.~Dufanets.
Morse ~fluids ~in the ~immediate ~vicinity of the critical point:
Calculation of thermodynamic coefficients.
\textit{J. Mol. Liq.} {\bf 385}, 122322 (2023),
https://doi.org/10.1016/j.molliq.2023.122322.

\bibitem{okumura_00} H.~Okumura, F.~Yonezawa.
Liquid-vapor coexistence curves of several interatomic model potentials.
\textit{J. Chem. Phys.} {\bf 113}, 9162 (2000),
https://doi.org/10.1063/1.1320828.

\bibitem{singh} J.K.~Singh, J.~Adhikari, S.K.~Kwak.
Vapor-liquid phase coe\-xis\-ten\-ce ~curves for ~Morse ~fluids.
\textit{Fluid Phase Equilib.} {\bf 248}, 1 (2006),
https://doi.org/10.1016/j.fluid.2006.07.010.

\bibitem{apf_11} E.M.~Apfelbaum.
The calculation of vapor-liquid coexistence curve of Morse fluid:
Application to iron. \textit{J. Chem. Phys.} {\bf 134}, 194506 (2011),
https://doi.org/10.1063/1.3590201.

\bibitem{martinez} A.~Mart\'{i}nez-Valencia, ~M.~Gonz\'{a}lez-Melchor,
~P.~Orea, ~J.~L\'{o}pez-Lemus.
~Liquid-vapour ~interface ~varying the softness and ran\-ge of
the ~interaction ~potential. ~\textit{Mol. Simul.} {\bf 39}, ~64 (2013),
https://doi.org/10.1080/08927022.2012.702422.

\bibitem{kpd118} M.P.~Kozlovskii, I.V.~Pylyuk, O.A.~Dobush.
The equation of state of a cell fluid model in the supercritical region.
\textit{Condens. Matter Phys.} {\bf 21}, 43502 (2018),
https://doi.org/10.5488/CMP.21.43502.

\bibitem{p120} I.V.~Pylyuk. Fluid critical behavior at liquid--gas
phase transition: Analytic method for microscopic description.
\textit{J. Mol. Liq.} {\bf 310}, 112933 (2020),
https://doi.org/10.1016/j.molliq.2020.112933.

\bibitem{rebenko_13}  A.L.~~Rebenko. ~~Cell ~~gas ~~model ~of ~classical
~statistical ~systems. ~~~\textit{Rev. Math. Phys.} ~~~{\bf 25}, ~~~1330006 ~~~(2013),
https://doi.org/10.1142/S0129055X13300069.

\bibitem{rebenko_15} V.A.~Boluh, A.L.~Rebenko. ~Cell gas free energy as
an approximation of the continuous model. ~\textit{J. Mod. Phys.} {\bf 6}, 168 (2015),
https://doi.org/10.4236/jmp.2015.62022.

\bibitem{pd120} I.V.~Pylyuk, O.A.~Dobush.
Equation of state of a cell fluid model with allowance for Gaussian
fluctuations of the order parameter. \textit{Ukr. J. Phys.} {\bf 65},
1080 (2020),
https://doi.org/10.15407/ujpe65.12.1080.

\bibitem{pk722} I.V.~Pylyuk, M.P.~Kozlovskii. First-order phase
transition in the framework of the cell fluid model: Regions of
chemical potential variation and the corresponding densities.
\textit{Ukr. J. Phys.} {\bf 67}, 54 (2022),
https://doi.org/10.15407/ujpe67.1.54.

\bibitem{ymo287} I.R.~Yukhnovskii. \textit{Phase Transitions of the
Second Order. Collective Variables Method} (World Scientific, 1987),
ISBN-10: 9971500876, ISBN-13: 9789971500870.

\bibitem{YuKP_2001} I.R.~Yukhnovskii, M.P.~Kozlovskii, I.V.~ Pylyuk.
\textit{Microscopic Theory of Phase Transitions in the Three-Dimensional Systems}
(Eurosvit, 2001) [in Ukrainian],
ISBN: 966-7343-26-X.

\bibitem{ykp202} I.R.~Yukhnovskii, M.P.~~Kozlovskii, I.V.~~Pylyuk.
~Thermodynamics of three-dimensional ~Ising-like ~systems ~in the higher
non-Gaussian ~approximation: ~Calculational ~method ~and ~dependence on
microscopic parameters. \textit{Phys. Rev. B} {\bf 66}, 134410 (2002),
https://doi.org/10.1103/PhysRevB.66.134410.

\bibitem{kpp406} M.P.~Kozlovskii, I.V.~Pylyuk, O.O~Prytula. Microscopic
~description of the critical behavior of three-dimensional ~Ising-like
~systems in an external field. ~\textit{Phys. Rev. B} ~{\bf 73},
~174406 (2006),
https://doi.org/10.1103/PhysRevB.73.174406.

\bibitem{kpp606} M.P.~Kozlovskii, I.V.~Pylyuk, O.O~Prytula. Free energy and
equation of state of Ising-like magnet near the critical point.
\textit{Nucl. Phys. B} {\bf 753}, 242 (2006),
https://doi.org/10.1016/j.nuclphysb.2006.07.006.

\bibitem{hensel} F.~Hensel. Critical behaviour of metallic liquids.
\textit{J. Phys.: Condens. Matter} {\bf 2}, SA33 (1990),
http://iopscience.iop.org/0953-8984/2/S/004.

\bibitem{llmo176} L.D.~Landau, E.M.~Lifshitz. \textit{Statistical Physics, Part 1}
(Nauka, 1976) [in Russian],
ISBN: 5922100548.

\bibitem{lgmo177} M.E.~Lines, A.M.~Glass. \textit{Principles and Application
of Ferroelectrics and Related Materials} (Clarendon Press, 1977),
ISBN-10: 0198512864, ISBN-13: 9780198512868.

\bibitem{wk174} K.G.~Wilson, J.~Kogut. The renormalization group and
the $\epsilon$ expansion. \textit{Phys. Rep.} {\bf 12}, 75 (1974),
https://doi.org/10.1016/0370-1573(74)90023-4.

\bibitem{ekd196} C.A.~Eckert, B.L.~Knutson, P.G.~Debenedetti. Supercritical
fluids as solvents for chemical and materials processing.
\textit{Nature} {\bf 383}, 313 (1996),
https://doi.org/10.1038/383313a0.
\end{thebibliography}
\end{document}